\documentstyle[aps, multicol]{revtex}

\def\etal{\emph{et al.}}

\input BoxedEPS
\SetepsfEPSFSpecial
\HideDisplacementBoxes

\begin{document}
\draft

\title{Proximity effect thermometer for local temperature measurements
on mesoscopic samples}
\author{J.\ Aumentado,$^1$ J. Eom,$^2$ V. Chandrasekhar,$^1$ P.M.
Baldo,$^3$ and L.E. Rehn$^3$}
\address{$^1$Department of Physics and Astronomy, Northwestern
University, Evanston, IL 60208}

\address{$^2$The James Franck Institute and Department of Physics,
University of Chicago, 5640 S. Ellis Avenue, Chicago, IL 60637} 
\address{$^3$Materials Science Division, Argonne National Laboratory,
Argonne, IL 60439}

\date{April 16, 1999}

\maketitle

\begin{abstract}  Using the strong temperature dependent resistance of
a normal metal wire in proximity to a superconductor, we have been able
to measure the local temperature of electrons heated by flowing a dc
current in a metallic wire to within a few tens of millikelvin at low
temperatures.  By placing two such thermometers at different parts of a
sample, we have been able to measure the temperature difference induced
by a dc current flowing in the sample.  This technique may provide a
flexible means of making quantitative thermal and thermoelectric
measurements on mesoscopic metallic samples.      
\end{abstract}

\pacs{74.80.Fp, 74.25.Fy, 74.80.-g, 74.80.Dm}

\begin{multicols}{2}
For many experiments on mesoscopic metallic samples, the issue of
determining the effective temperature $T_e$ of the electrons at low
temperature is of critical importance.  This is because it is $T_e$
which determines the electronic properties of the system, and not the
temperature $T_b$ of the thermal bath in which the sample is placed. 
With an energy input into the electron gas, the difference between
$T_e$ and $T_b$ can be large, particularly at low temperatures, where
the rapid decrease in the electron-phonon interaction means that the
electron gas is out of equilibrium with the phonon bath.  Hence it is
not possible to determine the electron temperature by  conventional low
temperature thermometers (which are typically well coupled only to the
phonon bath), and the need arises for thermometers which directly
measure the electron temperature.

In samples whose dimensions are much shorter than the electron-electron
scattering length
$L_{ee}$, the electron system itself may not be in equilibrium, so that
the question of the effective electron temperature is not a valid one. 
This was elegantly demonstrated in a recent experiment by Pothier
\etal \cite{pothier}, who measured the nonequilibrium electron
distribution function in a short metal wire using mesoscopic
superconductor-insulator-normal metal (SIN) junctions.  For samples
whose dimensions are longer than
$L_{ee}$ but shorter than the electron-phonon scattering length
$L_{ep}$, however, one can think about a position dependent local
electronic temperature $T_e$ which can be substantially different from
$T_b$ \cite{roukes}.  Previous techniques to measure $T_e$ have included
correlating the temperature with the Johnson noise measured across a
sample \cite{roukes,gifford,henny}, utilizing the temperature dependence
of the weak localization contribution to the magnetoresistance
\cite{mittal}, or measuring the current voltage characteristics of
a metallic system weakly coupled to a superconductor \cite{pothier}.

In this Letter, we describe a thermometer which makes use of the large
temperature dependent resistance of the superconducting proximity
effect to measure the temperature at different points of a complex
mesoscopic sample.  In contrast to techniques utilizing noise
measurements or weak localization, which only measure the average
temperature over relatively long samples,  this thermometer can measure
the electron temperature over size scales as small as $\sim 100$ nm. 
Consequently, one can determine the gradient of the electron
temperature in a mesoscopic sample, which may prove  useful in making
\emph{quantitative} thermal and thermoelectric measurements on
mesoscopic samples.

\begin{figure}[p]
\begin{center}
\BoxedEPSF{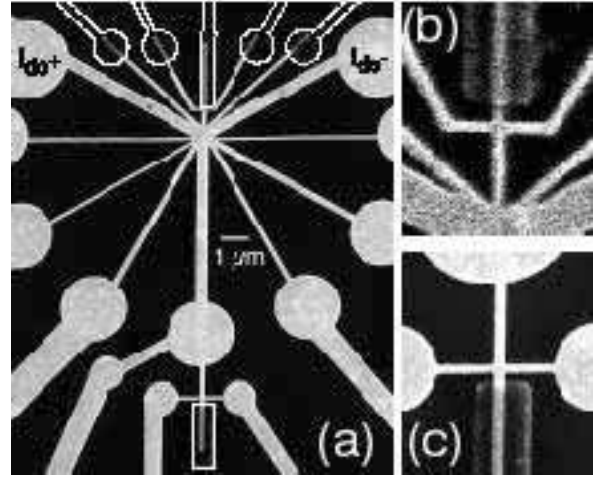 scaled 1000}
\end{center}
\caption{SEM images of device structure. (a) Large area layout. Al
structures are outlined with dashed lines for clarity. The wide V-shaped
wire with contacts marked $I_{dc} \pm$ which runs left to right at the
top is the heater. (b) Closeup view of top thermometer. (c) Closeup
view of bottom thermometer.}
\label{fig1}
\end{figure}

The samples for this experiment were fabricated using standard
multi-level electron beam lithography techniques.  Figure 1 shows a
scanning electron micrograph of one of our samples.  The design of the
samples was driven by our ongoing experiments on the width
dependence of the thermopower in mesoscopic Kondo wires \cite{eom},
which will be described in detail elsewhere.  The bright regions in the
micrograph of Fig. 1(a) correspond to a 51 nm thick Au film which was
deposited in the first level of lithography.  After implantation of
this entire layer with Fe ions to a concentration of $\sim 100$ ppm
\cite{implantnote}, a 75 nm thick Al film was deposited to form the
proximity effect thermometers (the Al film is outlined by dashed white
lines for clarity).  The sample consists of a number of wires which
meet at a central node.  The 0.36
$\mu$m wide V-shaped wire at the top of Fig. 1(a) is the heater wire
through which a dc current can be driven to raise the effective
temperature
$T_e$ of the electrons at the node above the bath temperature $T_b$. 
From the node, six wires of different widths radiate downward, each
terminating in a large circular pad.  These are the samples for the
width-dependent Kondo thermopower experiment, and they include a 0.35
$\mu$m wide control wire which runs vertically from the node to a
circular pad at the bottom of the micrograph.  

In this sample, there are two proximity effect thermometers which are
shown in greater detail in Figs. 1(b) and 1(c).  The first, which we
will denote the top thermometer (Fig. 1(b)), is attached to the top of
the node.  It consists of a
$\sim 0.5$
$\mu$m long Au\emph{Fe} wire with five terminals.  One terminal is
connected to an Al film which provides the superconducting reservoir
for the proximity effect.  The remaining contacts are used for making
four terminal resistance measurements on the wire.  These contacts are
connected to the wire bonding pads by superconducting Al lines.  Since
the superconductor has negligible thermal conductivity below its
transition temperature, this ensures that heat loss through these
contacts is minimized, and consequently, the thermal gradient across
the length of the thermometer is small.  The normal metal part of the
thermometer, which is evaporated at the same time as the heater,
is well coupled thermally to the node.  

The second thermometer, which we denote
the bottom thermometer (Fig. 1(c)), is attached to the circular
terminal pad of the wide control wire.  This thermometer is
$\sim1.0$ $\mu$m long, and can also be measured using four terminals,
two of which are connected through the circular pad.  One dangling lead
connected to the thermometer terminates in an Al film, which provides
the superconducting reservoir for the proximity effect for this
thermometer.  Unlike the top thermometer, connections to the wire
bonding pads are not made by superconducting leads, since the lower
thermometer is relatively far away from the heating source, and we
initially expected the temperature gradients at the pad to be small. 
It should be noted that for both thermometers, the four-terminal
resistance does not include the Al film, but only the proximity coupled
normal metal.

The proximity effect thermometers were first calibrated against a
RuO$_2$ thermometer attached to the mixing chamber of a dilution
refrigerator by measuring the resistance of the thermometers as a
function of temperature with no dc current flowing through the sample. 
\begin{figure}[p]
\begin{center}
\BoxedEPSF{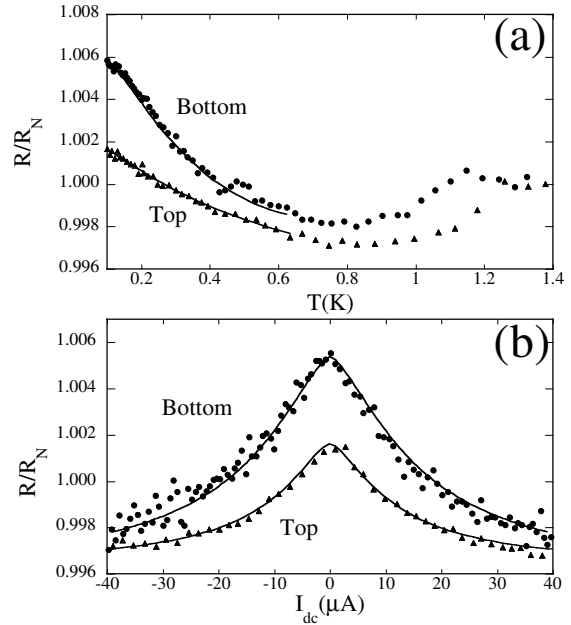 scaled 1000}
\end{center}
\caption{(a)  Resistance of top and bottom thermometers (normalized to
the resistance $R_N$ at $T=1.2$ K) as a function of temperature.  Solid
lines are fits to Eq.\ (\ref{RofT}) over the range 0.1 to 0.7 K.  (b) 
Normalized differential resistance of top and bottom thermometer as a
function of dc current through heater wire. Solid lines
are fits to Eq.\ (\ref{RofI}) over the range $I_{dc}=\pm 40$ $\mu A$,
with the parameters $a=2.1458, b=0.46022, c=40.348$ for the top
thermometer, and
$a=4.3007, b=2.5575, c=62.362$ (with appropriate units) for the bottom
thermometer. 
$R_N=2.1537$
$\Omega$ for the top thermometer, and 4.3185 $\Omega$ for the bottom
thermometer.}
\label{fig2}
\end{figure}
The resistance of the thermometers was measured using a homemade
four-terminal low-frequency ac resistance bridge with an excitation
current of
$\sim$500 nA, small enough to avoid self-heating, but large enough to
measure without averaging for very long times.  Figure 2(a) shows the
normalized resistance
$R/R_N$ of both thermometers as a function of the temperature $T$ as
measured by the RuO$_2$ thermometer.  As $T$ is reduced, a drop
in $R(T)$ is observed in both thermometers just below the transition
temperature
$T_c \sim 1.18$ K of the Al film.  As $T$ is lowered still
further,  the resistance increases, resulting in a minimum in $R(T)$
for both thermometers at
$T\sim0.8$ K.  This nonmonotonic or reentrant behavior of $R(T)$
in mesoscopic proximity coupled normal metals is well-known
\cite{charlat}, and is a result of the spatial dependence of the
electronic diffusion coefficient in the normal metal induced by its
proximity to the superconductor.  The overall change in resistance due
to the proximity effect in our samples is
$\sim 0.5-1$ percent, which is about a factor of 10 less than the
expected resistance change for reentrant behavior in a nonmagnetic metal
\cite{stoof}.  We believe that this reduction is due to the presence of
the implanted magnetic Fe impurities, which are expected to reduce the
electron phase coherence length in the normal metal, and hence the
amplitude of the proximity effect.  Nevertheless, the magnitude of the
resistance change is quite large.  For example, the resistance
contribution due to weak localization is typically on the order of
$10^{-4}$ of the total resistance \cite{altshuler}, two orders of
magnitude smaller than the temperature dependent resistance shown in
Fig. 2.  

In order to demonstrate the use of these thermometers to measure the
local temperature of the electrons in the Au\emph{Fe} film, we flow a dc
current $I$ through the wide heater strip using the terminals marked
$I_{dc}\pm$ in Fig. 1(a), while simultaneously measuring the resistance
of both the top and bottom thermometers with an ac resistance bridge. 
The dc current heats the electrons in the current path to a temperature
above
$T_b$.  The electrons are cooled through interaction with phonons in
the metal, and by electronic thermal conduction through the metallic
parts of the sample itself, which is more efficient near the large
contact pads of the sample.  This leads to a nonuniform electron
temperature profile in the heater wire, with the maximum electron
temperature at the node \cite{nagaev}.  Parts
of the sample which are connected to the heater but do not have a dc
current flowing through them (such as the control wire) will also
develop a temperature gradient as a function of the dc current through
the heater.   

Figure 2(b) shows the four terminal ac resistance $R$ of the top and
bottom thermometer as a function of $I$.  During this measurement, the
mixing chamber of the dilution refrigerator was maintained at $\sim
97.5$ mK.  Both curves are symmetric with respect to $I$, as
would be expected since the heating of the electron gas by the dc
current should be independent of the direction of the current.  By
correlating $R(I)$ shown in Fig.
2(b) with $R(T)$ shown in 
Fig. 2(a) for each thermometer, one can determine the effective
electronic temperature at the node and at the contact pad of the
control wire as a function of $I$.  In order to do this, we fit $R(T)$
for each thermometer to a fourth order polynomial of the form
\begin{equation}
R(T)=\sum_{j=0}^4 a_j (\text{log}T)^j
\label{RofT}
\end{equation}
over the temperature range 0.07-0.625 K.  The resulting fits are shown
as the solid lines in Fig. 2(a).  Similarly, $R(I)$ for
both thermometers were fit by an equation of the
form\cite{fitnote}
\begin{equation}
R(I)=a + \frac{b}{\left|I\right|^{3/2} + c}
\label{RofI}
\end{equation}
as demonstrated by the solid lines in Fig. 2(b), which are fits to
Eq.\ (\ref{RofI}) over the range $I_{dc}=\pm 40$ $\mu$A.  Finally,
the effective electron temperature
$T_e$ as a function of
$I$ is obtained by cross interpolating between $R(T)$ and $R(I)$
obtained from the fits for each thermometer\cite{expnote}.  Figure 3
shows $T_e (I)$ obtained in this manner for both thermometers.  This
plot indicates that even relatively small dc currents can substantially
raise the effective temperature
$T_e$ over the bath temperature $T_b$.  
\begin{figure}[p]
\begin{center}
\BoxedEPSF{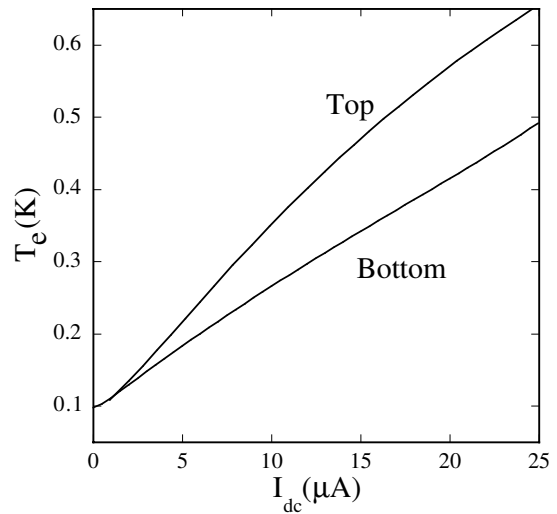 scaled 1000}
\end{center}
\caption{$T_e$ as a function of $I_{dc}$ obtained by cross interpolating
the fits for $R(T)$ and $R(I)$ shown in Fig. 2 for the top and bottom
thermometer.}
\label{fig3}
\end{figure}
For example, at a dc current of
$I\sim 5$ $\mu A$,
$T_e$ at the node is $\sim 218$ mK, an increase of $\sim$ 120 mK over
the bath temperature of 97.5 mK.  The surprising fact is that the
electron temperature at the lower thermometer also increases
substantially, to a value of 183 mK, giving a temperature differential
of $\sim$ 35 mK across the control wire.   At lower temperatures, where
cooling by phonons is less efficient, the heating effect would be
expected to be even more drastic.  These results underline the
importance of using low excitation currents in low temperature
transport measurements on mesoscopic samples in order to avoid
self-heating of the electrons.

The use of Al as the superconductor in these thermometers restricts
the temperature range of the thermometers to below 0.8 K, but this
range can easily be increased by using a superconductor with a higher
$T_c$ such as Pb or Nb.  The ability to measure the spatial variation
of the electron temperature that we have demonstrated here opens up the
possibility of using these thermometers to make quantitative thermal and
thermoelectric measurements on mesoscopic samples.

Support for this work was provided by the NSF through
grant number DMR-9801982, by the David and Lucile Packard
Foundation, and by the DOE-BES through contract number W-31-109-Eng-38. 

%-----------References---------------------------------------

%------------FIGURE CAPTIONS---------------------------------

\end{multicols}


\begin{references}
\bibitem{implantnote}The samples were implanted at doses of
4x$10^{12}$/cm$^2$ and 4x$10^{13}$/cm$^2$ at energies of 40 and 160 keV
respectively.
\bibitem{kondo}The resistance of our thermometers rises above the
normal state resistance, unlike the samples in Ref.
\cite{charlat}.  The reason for this is unclear at present, but it may
be due to repulsive electron-electron interactions in the normal metal.
\bibitem{fitnote}These functional forms (Eqs.\ (\ref{RofT}) and
(\ref{RofI})) were empirically determined to best describe the data with
the fewest number of fitting parameters.
\bibitem{expnote}For the bottom thermometer, the experimental $R(I)$ was
offset by 0.002 $\Omega$ to match the zero bias resistance with
the measured temperature dependent resistance at 97.5 mK before
performing the analysis in order to account for an offset in the
measurement.
\end{references}
\end{document}